# HIGH GRADIENT TESTS OF THE FERMILAB SSR1 CAVITY*

T. Khabiboulline#, C. M. Ginsburg, I. Gonin, R. Madrak, O. Melnychuk, J. Ozelis, Y. Pischalnikov, L. Ristori, A. Rowe, D. A. Sergatskov, A. Sukhanov, I. Terechkine, R. Wagner, R. Webber, and V. Yakovlev, FNAL, Batavia, IL 60510, USA

*Abstract*

In Fermilab we are building and testing superconducting Single Spoke Resonators (SSR1, β=0.22) which can be used for acceleration of low beta ions. The first two cavities performed very well during the cold tests in the Vertical Test Stand (VTS) at FNAL. One dressed cavity was also tested successfully in the Horizontal Test Stand. Currently we are building an eight-cavity cryomodule for PXIE project. An additional ten cavities were manufactured in industry and ongoing cold test results are presented in this paper.

## INTRODUCTION

Initially SSR1 was developed at Fermilab for acceleration of proton/H⁻ beams in the 10-30 MeV part of the HINS pulsed accelerator. The operating temperature of the cavity was 4.4K [1]. Currently Fermilab is designing and building Project X Injector Experiment (PXIE), a prototype of the Project X CW front end that will be used to prove the design concept. The CW operation and 2K operating temperature imply a change to the operating requirements for the SSR1 cavity [2].

## RF DESIGN OF SSR1 CAVITY

When we start designing of Spoke cavities accelerating gradient was calculated using effective accelerating length $L_{eff}$ based on ANL definition $L_{eff}=2/3\beta\lambda$, where λ is the wavelength at operating frequency. Previous cold test results were presented based on that definition. Most people are using another definition accelerating length $L_{eff}=\beta*\lambda/2*n$, where n is number of accelerating gaps in the cavity. To avoid further confusing it was decided to switch to this common definition [3]. All cold test results were rescaled to $L_{eff}=\beta\lambda$ in this paper. Summary of parameters for SSR1 is shown in the Table 1.

## VTS TESTS OF THE CAVITY S1-ZN-101

The 1st SSR1 resonator, S1-ZN-101, was fabricated by E. Zanon. The BCP was done at ANL using the standard HF:HNO3:H3PO4(1:1:2) acid mixture [4]. During BCP, the cavity was oriented with the power coupler port and the vacuum port along the vertical axis and the beam pipes along the horizontal axis.

In order to obtain uniform etching of ~120 μm, the cavity was flipped top to bottom between the two etching sessions. After BCP, the S1-ZN-101 was moved to the class 10 clean area for HPR. After completing the HPR, the cavity was left in a good orientation for drainage (vacuum port up) and left to dry in the class 10 clean area overnight. Blank-off flanges were installed the next day, and the cavity was taken to the class 10 clean room in the MP9 building at Fermilab for final preparations before testing in the VTS.

Table 1: SSR1 cavity parameters for Project X.

| | |
|---|---|
| Operating frequency | 325 MHz |
| Optimum beta, β | 0.22 |
| $L_{eff}=2\beta(\lambda/2)=\beta\lambda$ | 203 mm |
| Accelerating gradient, $E_{acc}$ | 12 MV/m |
| $E_{peak}/E_{acc}$ | 3.84 |
| $B_{peak}/E_{acc}$ | 5.81 mT/(MV/m) |
| $Q_0$ | $5*10^9$ |
| G factor | 84 Ω |
| R/Q | 242 Ω |

There have been several test sessions in the VTS without Hydrogen degassing procedure. For the first cool down VTS did not yet have a cavity vacuum system, so the SSR1 was evacuated (to $2\times10^{-6}$ Torr) and sealed before installation into the Dewar. A residual surface resistance of $R_0$ = 5.1 nΩ was measured.

2nd cool down began with a better warm cavity vacuum of $1.2\times10^{-7}$ Torr, due to active pumping. At 2 K $E_{acc}$, reached 9 MV/m before field emission prevented further increase with our 200 W power supply, Fig. 1. Total multipactor processing time was ~10 hours.

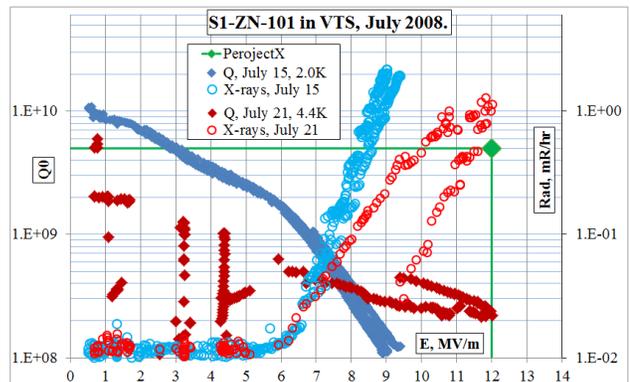

Figure 1: VTS test summary of the cavity S1-ZN-101.

X-ray detector is installed just above the VTS Dewar's top plate. Green diamond corresponds to design parameter of SSR1 cavity for the Project X. In the latest cold test cavity was tested only at 4.4K temperature and $E_{acc}$ reached 12 MV/m after 2 hours of power processing. Superfluid helium leak developed in the pumping line after pumping down. At the moment 4.4K was design operating temperature and due to scheduling 2K test was skipped.

## VTS TESTS OF THE CAVITY S1-RK-102

Second SSR1 cavity S1-RK-102 was manufactured in ROARK under supervision of Fermilab. Chemical treatment was similar to S1-ZN-101. Because of 600C 10 hours heat treatment for Hydrogen degassing, the 1$^{st}$ cool down of S1-RK-102 was free of the effects of Q disease. The effect of Q disease at 4.4 K was not very pronounced with the usual fast cool down of the VTS [5].

As usual for SSR1 cavities, multipactor processing take place at low accelerating gradients 1-7 MV/m in the beginning of the test. When accelerating gradient reached 8 MV/m multipactor processing done and accelerating gradient jumped to 16 MV/m limited by thermal quench which was induced by field emission. Gradient jumped again to 22 MV/m, after additional multipactor processing, limited by thermal quench. X-ray radiation also dropped by factor of two orders. At 4K 16.5 MV/m limited by available power, but very close to thermal quench limit, Fig. 2. In the temperature range 2.2-4.4K accelerating gradient was limited also by thermal quench.

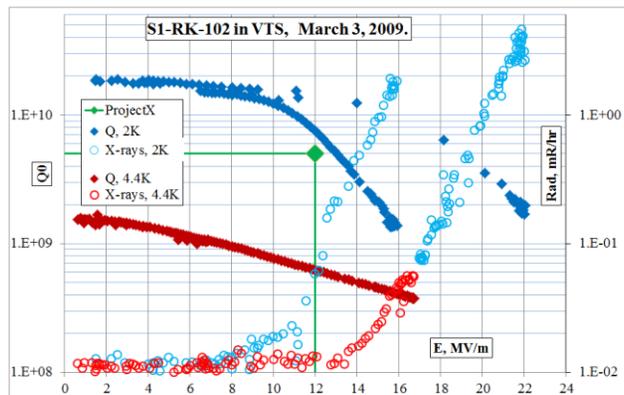

Figure 2: VTS test summary of S1-RK-102.

At 2 K, the large slope had disappeared and the cavity reached Eacc = 22 MV/m. After warming up to 4.4 K, a field of 16 MV/m was achieved. A residual surface resistance of $R_{res}$ = 4.8 nΩ was measured.

## VTS TESTS OF THE CAVITY S1-NR-105

S1-NR-105 is the first cavity from industrial production set of ten cavities manufactured by Niowave - Roark collaboration. BCP and HPR of this cavity were done in the new ANL-FNAL facility. Positioning of the cavity in HPR was different from previous two cavities. Before final BCP cavity was heat treated at 600C for 10 hours in the new FNAL baking furnace. Cold test in VTS1 started with severe multipactoring. After long ~ 60 hours of power processing accelerating gradient reached 11.5 MV/m, Fig. 3.

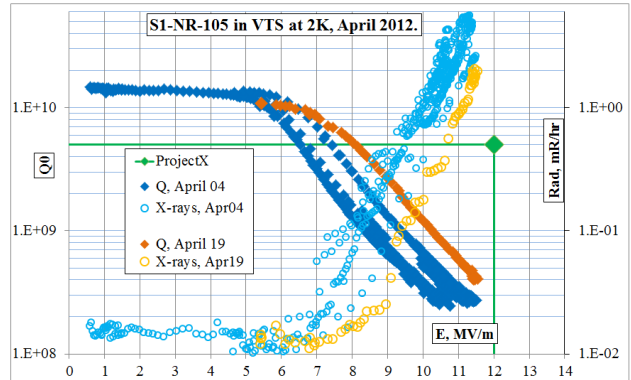

Figure 3: 1$^{st}$ cold test results of the cavity S1-NR-105.

It was suspected that more severe multipactoring was related with extra water in the cavity volume. To remove water from the cavity additional 120C 48 hours baking was implemented. Baking helped to improve $Q_0$ by factor of two. Maximum accelerating gradient was improved very little, less than 5%.

## HTS TESTS OF THE CAVITY S1-ZN-101

The jacketed S1-ZN-01 cavity was baked at 600C 2hours and dressed with stainless steel Helium jacket. With prototype tuners and a high, ~1.5e8, $Q_{ext}$ drive antenna for CW testing in Spoke Cavity Test Facility SCTF [6], Fig. 4. Upon initial cool down to ~4.5 K, the cavity resonant frequency was within 20 kHz of the expected value. The cavity and stainless steel vessel behaved as expected during cool down [7].

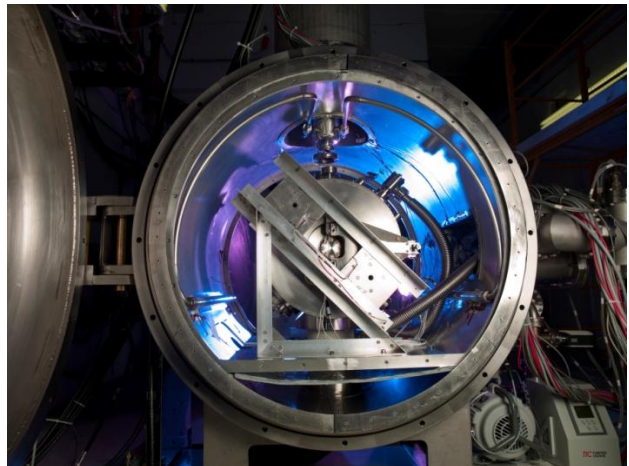

Figure 4: S1-ZN-101 with helium vessel inside of the 1.44 m OD SCTF cryostat.

The cavity tuners [8] were engaged to apply an inward force of 6 kN on the beam pipe on each end of the cavity. In this state, the resonant frequency sensitivity to helium pressure was measured to be -145 ± 15 Hz/Torr. The static Lorentz force detuning coefficient was found to be -1.5 ± 0.5 Hz/(MV/m)$^2$.

RF conditioning to clean up multipactoring and attain high field operation required ~10 hours. The ability to switch the CW RF amplitude between relatively low and high power with a variable duty cycle and a few second period proved useful. Operating in this mode was necessary for high $E_{acc}$ operation, where x-ray emissions increase power loss. Drive power was switched between low and high values with suitable duty factor to keep the average loss below the 35 watt limit imposed by system cooling and pressure constraints. Maximum accelerating gradient 18 MV/m was reached, x-rays started above 14 MV/m, Fig. 5.

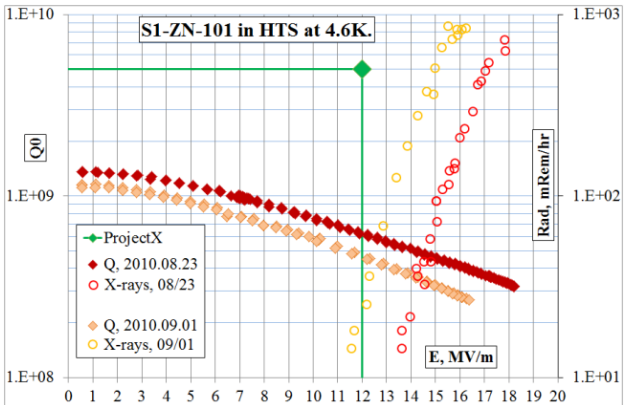

Figure 5: $Q_0$ vs. $E_{acc}$ and x-ray activity of S1-NR-101 in SCTF. Test on September 1$^{st}$ done after slow cool down.

## PULSE TESTS OF THE CAVITY S1-NR-101

In pulsed test mode, the cavity is driven by a 2.5 MW klystron. Klystron output is through WR2300 waveguide, a nominally 10 dB waveguide coupler and 3 inch coaxial waveguide. During cavity testing, power is further limited to ~ 65 kW by attenuators on the klystron RF input. This is sufficient to drive the cavity to its maximum field in approximately 1 ms. Note that since $Q_{ext}$ ~10$^6$ and $Q_0$ ~10$^9$, most of the power is reflected [9]. The accelerating gradient was increased to 21 and then 24 MV/m, during which quenching and large X-ray bursts were observed. After running the cavity for approximately an hour at 24-25 MV/m, the X-ray bursts eventually stopped and the quiescent X-ray level also decreased, Fig.6. This suggests that a field emitter had been processed away.

In the SCTF, the cavity and helium vessel are cooled to 4.5 K in a test cryostat [9]. The test cryostat contains a warm magnetic shield and an 80 K thermal shield.

Resonance frequency tracking is necessary because of significant pressure fluctuations of ~12 Torr and cavity frequency sensitivity to the pressure fluctuations was 145 Hz/Torr. The PLL has a bandwidth of +/-20 kHz. A sample-and-hold approach was implemented to track the cavity resonance during the pulse.

Fast tuners and associated algorithms for microphonics compensation and Lorentz force detuning compensation study were done as part of this test program.

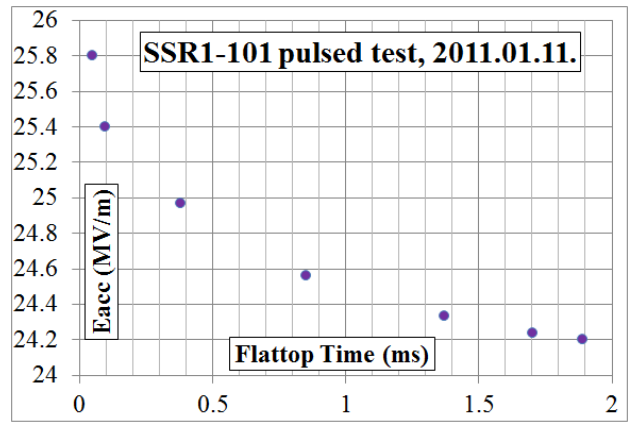

Figure 6: Accelerating gradient $E_{acc}$ vs. flattop time.

## SUMMARY

Two cavities out of three tested so far, S1-ZN-101 and S1-RK-102, reached Project X design gradient of 12 MV/m. Quality factor of the S1-RK-102 at 2K was also acceptable and above 5*10$^9$ design requirement at 12 MV/m. We are planning further testing of the jacketed cavity S1-ZN-101 at 2K after 2K upgrade of SCTF. Measured $Q_0$ of the cavity S1-NR-105 is very low due to severe field emission. This cavity will be tested again after additional HPR.